\documentclass[utf8]{frontiersSCNS} % for Science, Engineering and Humanities and Social Sciences articles
\usepackage{url,hyperref,lineno,microtype,subcaption}
\usepackage[onehalfspacing]{setspace}

\def\keyFont{\fontsize{8}{11}\helveticabold }
\def\firstAuthorLast{Fan {et~al.}} %use et al only if is more than 1 author
\def\Authors{Yuhong Fan\,$^{1,*}$ and Tie Liu\,$^{2,3}$ }
\def\Address{$^{1}$High Altitude Observatory, National Center for Atmospheric Research, Boulder, CO, USA \\
$^{2}$Key Laboratory for Dark Matter and Space Science, Purple Mountain Observatory, CAS, Nanjing, Jiangsu, China \\
$^{3}$School of Astronomy and Space Science, University of Science and Technology of China, Hefei, Anhui, China}
\def\corrAuthor{Yuhong Fan}

\def\corrEmail{yfan@ucar.edu}

\begin{document}
\onecolumn
\firstpage{1}

\title[Prominence-cavity system]{MHD simulation of prominence-cavity system} 

\author[\firstAuthorLast ]{\Authors} %This field will be automatically populated
\address{} %This field will be automatically populated
\correspondance{} %This field will be automatically populated

\extraAuth{}% If there are more than 1 corresponding author, comment this line and uncomment the next one.
%\extraAuth{corresponding Author2 \\ Laboratory X2, Institute X2, Department X2, Organization X2, Street X2, City X2 , State XX2 (only USA, Canada and Australia), Zip Code2, X2 Country X2, email2@uni2.edu}

\maketitle

\begin{abstract}

%%% Leave the Abstract empty if your article does not require one, please see the Summary Table for full details.
\section{}
We present magnetohydrodynamic simulation of the evolution
from quasi-equilibrium to onset of eruption
of a twisted, prominence-forming coronal magnetic flux rope underlying
a corona streamer.
The flux rope is built up by an imposed flux emergence
at the lower boundary.
During the quasi-static phase of the evolution, we find the formation
of a prominence-cavity system with qualitative features resembling
observations, as shown by the synthetic SDO/AIA EUV
images with the flux rope observed above the limb viewed nearly along
its axis.  The cavity contains substructures including
``U''-shaped or horn-like features extending from the prominence enclosing
a central ``cavity'' on top of the prominence.
The prominence condensations form in the dips of the highly twisted field
lines due to runaway radiative cooling and the cavity is formed by the
density depleted portions of the prominence-carrying field lines extending
up from the dips.
The prominence ``horns'' are threaded by twisted field lines containing
shallow dips, where the prominence condensations have evaporated to
coronal temperatues.
The central ``cavity'' enclosed by the horns is found to correspond
to a central hot and dense core containing twisted field lines that
do not have dips.
The flux rope eventually erupts as its central part rises quasi-statically
to a critical height consistent with the onset of the torus instability.
The erupting flux rope accelerates to a fast speed of nearly 900 km/s
and the associated prominence eruption shows significant rotational
motion and a kinked morphology.

\tiny
 \keyFont{ \section{Keywords:} magnetohydrodynamics (MHD), methods: numerical simulation, Sun:corona, Sun:coronal mass ejection, Sun: magnetic fields, Sun: prominences } %All article types: you may provide up to 8 keywords; at least 5 are mandatory.
\end{abstract}

\section{Introduction}

Solar filaments and prominences are observed to be a major precursor
of coronal mass ejections (CMEs) \citep[e.g.][]{Webb:Hundhausen:1987}.
When observed in white light or EUV
above the limb viewed nearly along their lengths, they often display a
prominence-cavity system with a relatively dark cavity surrounding the
lower central prominence \citep[see review by][]{Gibson:2015}.
EUV observations of prominence-cavity systems have also shown
substructures within the cavities with ``U''-shaped
prominence ``horns'' extending from the prominence, enclosing a
central ``cavity'' or ``void'' on top of the prominence,
see e.g. Figure 8 and Figure 12 in \citet{Gibson:2015}
and Figure 2(c) in \citet{Su:etal:2015}.
The first 3D MHD simulations of prominence formation in a stable
equilibrium coronal magnetic flux rope were carried out
by \citet{Xia:etal:2014,Xia:Keppens:2016a}.
With the use of adaptive grid refinement and including
the chromosphere as the lower boundary, their 3D simulations
obtained a prominence-cavity system with the prominence showing
fine-scale, highly dynamic fragments, reproducing
many observed features seen in SDO/AIA observations.

Recently, \citet{Fan:2017} (hereafter F17) and \citet{Fan:2018} (hereafter F18)
have carried out 3D MHD simulations of prominence forming coronal flux ropes
under coronal streamers, with the flux rope evolving from
quasi-equilibrium to onset of eruption, leading to a CME with
associated prominence eruption.
In those simulations, a significantly twisted, longitudinally
extended flux rope is built up in the corona under a pre-existing
coronal streamer solution by an imposed flux emergence at the lower
boundary. During the quasi-static evolution of the emerged flux rope,
cool prominence condensations are found to form in the dips of the
significantly twisted field lines due to the radiative instability  
driven by the optically thin radiative cooling of the relatively
dense plasma in the emerged dips.
In the prominence-forming flux rope simulation in F18 (labeled as the
``PROM'' simulation in that paper), we find that the prominence weight
is dynamically important and can suppress the onset of the kink
instability and hold the flux rope in quasi-equilibrium for a
significantly longer period of time, compared to a case without
prominence formation.  We also find the formation of a cavity
surrounding the prominence, and substructures inside the cavity
such as prominence ``horns'' and a central ``cavity'' on top
of the prominence.  However in the simulations in
F18, a pre-existing streamer solution (the ``WS'' solution
in F17) with a wide mean foot-point separation of the arcade field
lines are used and the corresponding potential field has
a slow decline with height.  As a result, we obtained a prominence
and cavity that extend to rather large heights that are
larger than typically observed before the onset of eruption.
Here we extend the work of F17 and F18 by modeling the
prominence-forming flux rope under a pre-existing coronal
streamer with a significantly narrower mean foot-point separation
for its closed arcade field lines.
We find the formation of a prominence-cavity system with the heights
for the prominence and the cavity that are more in accordance with 
those of the typically observed quiescent prominence-cavity systems.
We carry out a more detailed analysis of the characteristics of
the 3D magnetic fields comprising the different features of the
prominence-cavity system.
We also find that the flux rope begins to erupt at a significantly
lower height, consistent with the onset of the torus instability, and
results in a fast CME with an associated prominence eruption that
shows a kinked morphology.

\section{Model Description}
\label{sec:model}

For the MHD simulation presented in this paper, we use the same numerical MHD model
described in detail in F17.  The readers are referred to that paper's
sections 2 and 3.1 for the description of the equations solved, the numerical code,
and the initial and boundary conditions for the simulation set-up. As a brief overview,
we use the ``Magnetic Flux Eruption'' (MFE) code to solve the set of
semi-relativistic MHD equations (eqs. (1)-(6) in F17) in spherical geometry,
with the energy equation explicitly taking into account the non-adiabatic effects
of an empirical coronal heating (which depends on height only), optically thin
radiative cooling, and the field-aligned electron heat conduction.
The inclusion of these non-adiabatic affects allow for the development
of the radiative instability that leads to the formation of prominence
condensations in the coronal flux rope as shown in the simulations of
F17 and F18.
The simulation domain is in the corona, ignoring the photosphere and chromosphere
layers, with the lower boundary temperature ($T=5 \times 10^5$ K) set at the
base of the corona, but with an adjustable base density (and hence base pressure)
that depends upon the downward heat conduction flux to crudely represent the
effect of chromospheric evaporation (equations (17) and (18) and the
associated descriptions in F17).
The radiative loss function $\Lambda (T)$ used for the radiative cooling in
equation (13) in F17 is the ``actual'' curve shown in Figure 1 of F17 (also
the same as the ``PROM'' case in F18). 
As described in F17, the radiative loss function used is modified to 
suppress cooling for $T \leq 7 \times 10^4$ K, so that the smallest pressure
scale height of the coolest plasma that can form does not go below two
grid points given our simulation resolution.
In the following we describe the specific changes that have been made in
the set-up of the current simulation.

The empirical coronal heating used in this simulation is modified to use
two exponentially decaying (with height) components instead of just one
used in F17, i.e.
we change equation (14) in F17 to the following:
\begin{equation}
H = \frac{F_1}{L_1} \frac{R^2_s}{r^2}
\exp \left[ -(r - R_s) / L_1 \right ]
+ \frac{F_2}{L_2} \frac{R^2_s}{r^2}
\exp \left[ -(r - R_s) / L_2 \right ] 
\label{eq:heatingfunc}
\end{equation}
where the input energy flux densities for the two components are
$F_1=F_2=5 \times 10^5 \, {\rm ergs} \, {\rm cm}^{-2} {\rm s}^{-1}$,
and the decay lengths are $L_1 = 5 \times 10^{10}$ cm and
$L_2 = 2.5 \times 10^9$ cm,
$r$ is the radial distance to the center
of the sun and $R_s$ denotes the solar radius.
The first much more extended component
is aimed to heat and accelerate the background solar wind and open up
the ambient coronal magnetic field. The second more spatially
confined heating is aimed to enhance the heating near the base to
enhance the base pressure and plasma inflow into the corona, which promotes the
formation of prominence condensations in the emerged flux rope.

\begin{figure}[ht!]
\begin{center}
\includegraphics[width=4in]{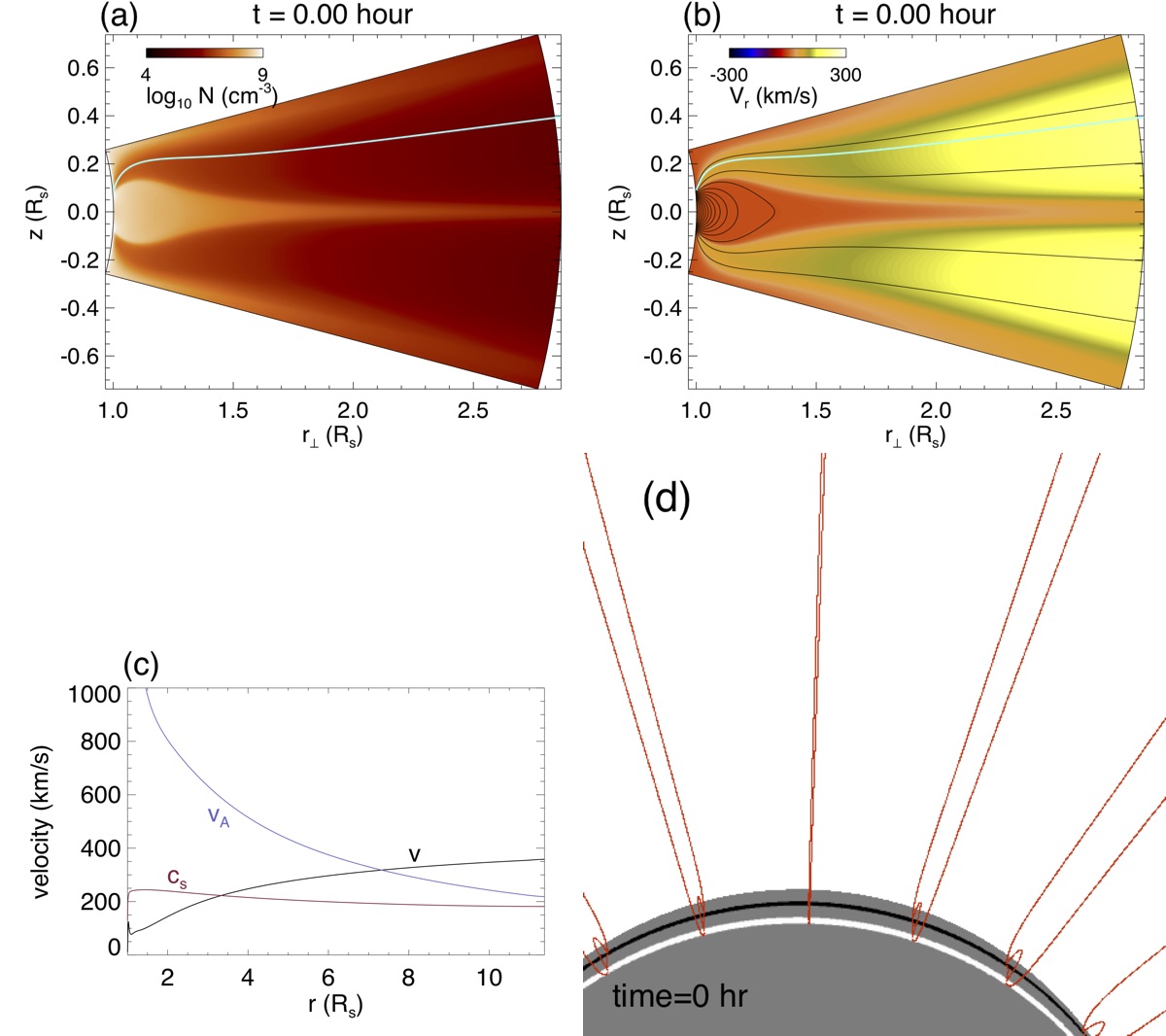}
\end{center}
\caption{The relaxed 2D quasi-steady streamer solution for the
initial state. (a) shows the cross-section density, (b) shows the
cross-section radial velocity
over-plotted with magnetic field lines,
(c) shows the parallel velocity V, the Alfv\'en
speed $V_A$, and sound speed $C_S$ along an open field line (marked
as the green line in (a) and (b)) in the ambient solar wind.
(d) shows a 3D view of the initial streamer field lines in the
simulation domain with the lower boundary color
showing the normal flux distribution of the initial bipolar bands.}
\label{fig:init_streamer}
\end{figure}

As described in section 3.1 of F17, we first initialize a 2D
quasi-steady solution of a coronal steamer with an ambient solar
wind in a longitudinally extended (in $\phi$) spherical wedge domain.
We use the same simulation domain (with $r \in [R_s, 11.47 R_s]$,
$\theta \in [75^{\circ}, 105^{\circ}]$,
and $\phi \in [-75^{\circ}, 75^{\circ}]$,
where $R_s$ is the solar radius) and the grid as those for
the ``WS-L''
(wide-streamer/long flux-rope)
simulation in F17 (also the ``PROM'' simulation in F18).
However, we use the initial normal flux distribution of a narrow bipolar
pair of bands on the lower boundary as that for the NS (narrow-streamer)
solution in F17 (see Figure 2(b) in F17), and increase the field
strength by a factor of two.
We obtain the relaxed 2D quasi-steady streamer solution for
the initial state as shown in
Figure \ref{fig:init_streamer}.
The cross-sections (panels (a) and (b)) of the initial state show
a dense helmet dome of closed magnetic field
approximately in static equilibrium, surrounded by an ambient open
field region with a solar wind outflow with flow speed that
accelerates to supersonic and super Alfv\'enic speed (see panel (c)).
A 3D view of the initial streamer field lines in the simulation
domain is shown in Figure \ref{fig:init_streamer}(d). 

Into this initial streamer field, we then impose at the lower
boundary the emergence of a twisted magnetic torus by specifying
an electric field as described in 
F17 (see equations (19)-(22) and the associate description in F17).
The specific parameters for the driving emerging torus
(see the definitions of the parameters in F17) used for the
present simulation are:
the minor radius $a=0.04314 R_s$,
twist rate per unit length $q/a = - 0.0166$ rad ${\rm Mm}^{-1}$,
major radius $R'=0.75 R_s$,
axial field strength $B_t a/R' = 106$ G, and
the driving emergence speed $v_0 = 1.95$ km/s.
The driving flux emergence at the lower boundary is stopped
when the total twist in the emerged flux rope reaches
about 1.76 winds of field-line twist between the two anchored
ends.

We note that the present simulation and the PROM simulation
in F18 are similar in the driving flux emergence,
where a long flux rope of similar total twist is driven into the
corona.
The essential difference is that the pre-existing arcade
field in the streamer of the present simulation has a
significantly narrower foot-point separation and a stronger
foot-point field strength (compare Figure \ref{fig:init_streamer}(b)
in this paper and Figure 2(b) in F18).
The mean foot-point separation
in the present case is $0.13 R_s$ compared to $0.24 R_s$ for that
in the PROM simulation in F18.
The narrower foot-point separation results in a
faster decline of field strength with height for the
corresponding potential field in the present case.
With a stronger arcade foot-point field strength, we expect a stronger
confinement of the emerging flux rope to a lower height, and hence
smaller cavity and prominence heights,
improving upon the PROM case in F18, in which the cavity and prominence
heights obtained are too high compared to the typical observed values.
Furthermore, the stronger field strength
lower down and the faster decline with height
of the corresponding potential field are expected to have significant
effects on the height for the onset of the torus instability and
the acceleration of the erupting flux rope as shown in previous
simulations by \citet{Toeroek:Kliem:2007}.
 
%For requirements for a specific article type please refer to the Article Types on any Frontiers journal page. Please also refer to  \href{http://home.frontiersin.org/about/author-guidelines#Sections}{Author Guidelines} for further information on how to organize your manuscript in the required sections or their equivalents for your field
% For Original Research articles, please note that the Material and Methods section can be placed in any of the following ways: before Results, before Discussion or after Discussion.

\section{Simulation Results}

\subsection{Overview of evolution}

\begin{figure}[ht!]
\begin{center}
\includegraphics[width=5in]{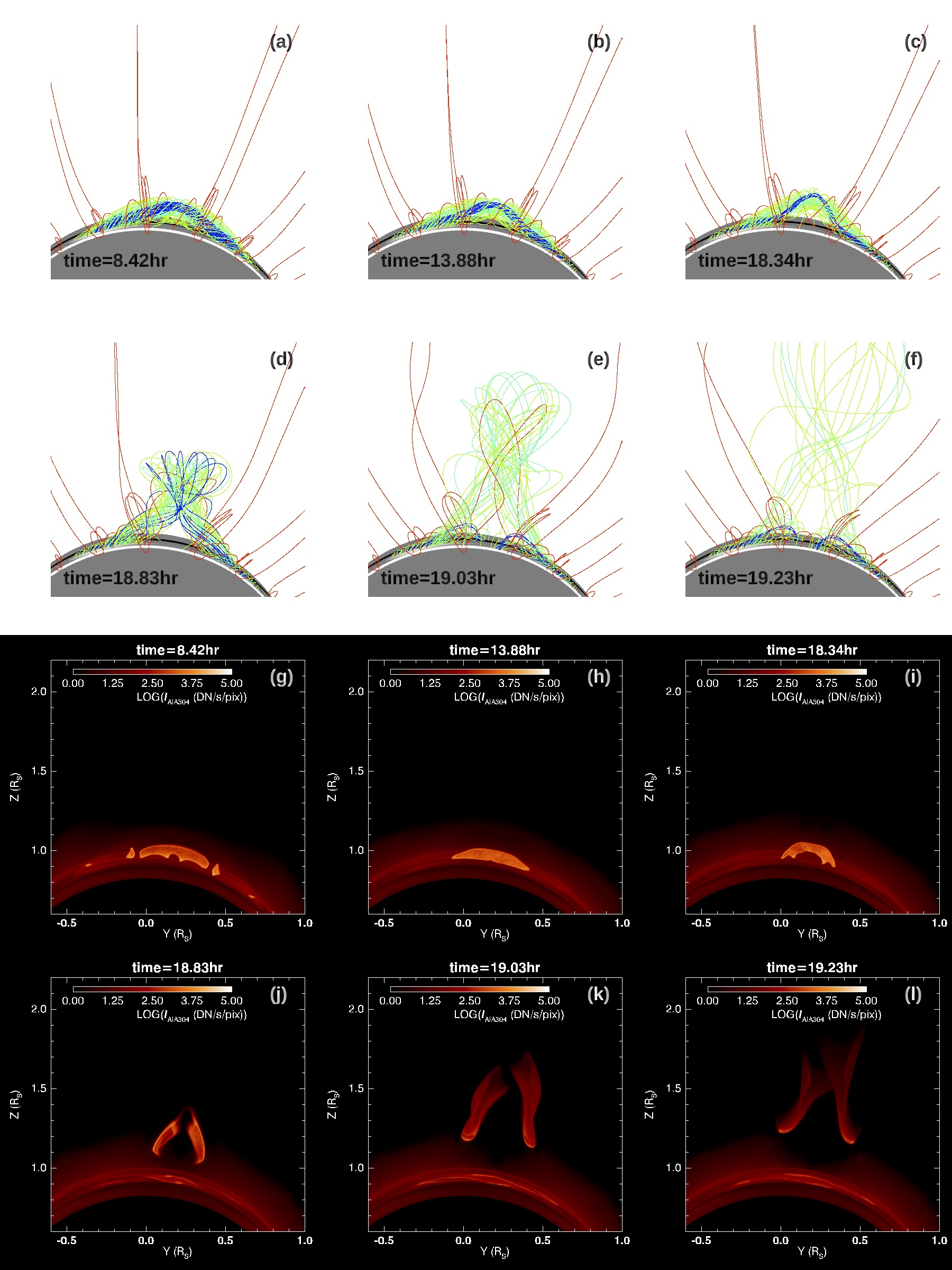}% This is a *.jpg file
\end{center}
\caption{(a)-(f) show a sequence of snapshots of the 3D magnetic field lines through the
course of the evolution of the emerged coronal flux rope, and (g)-(l) show the corresponding
synthetic SDO/AIA EUV images in 304 {\AA} channel from the same
perspective view.}
\label{fig:evol_sideview}
\end{figure}

\begin{figure}[ht!]
\begin{center}
\includegraphics[width=4in]{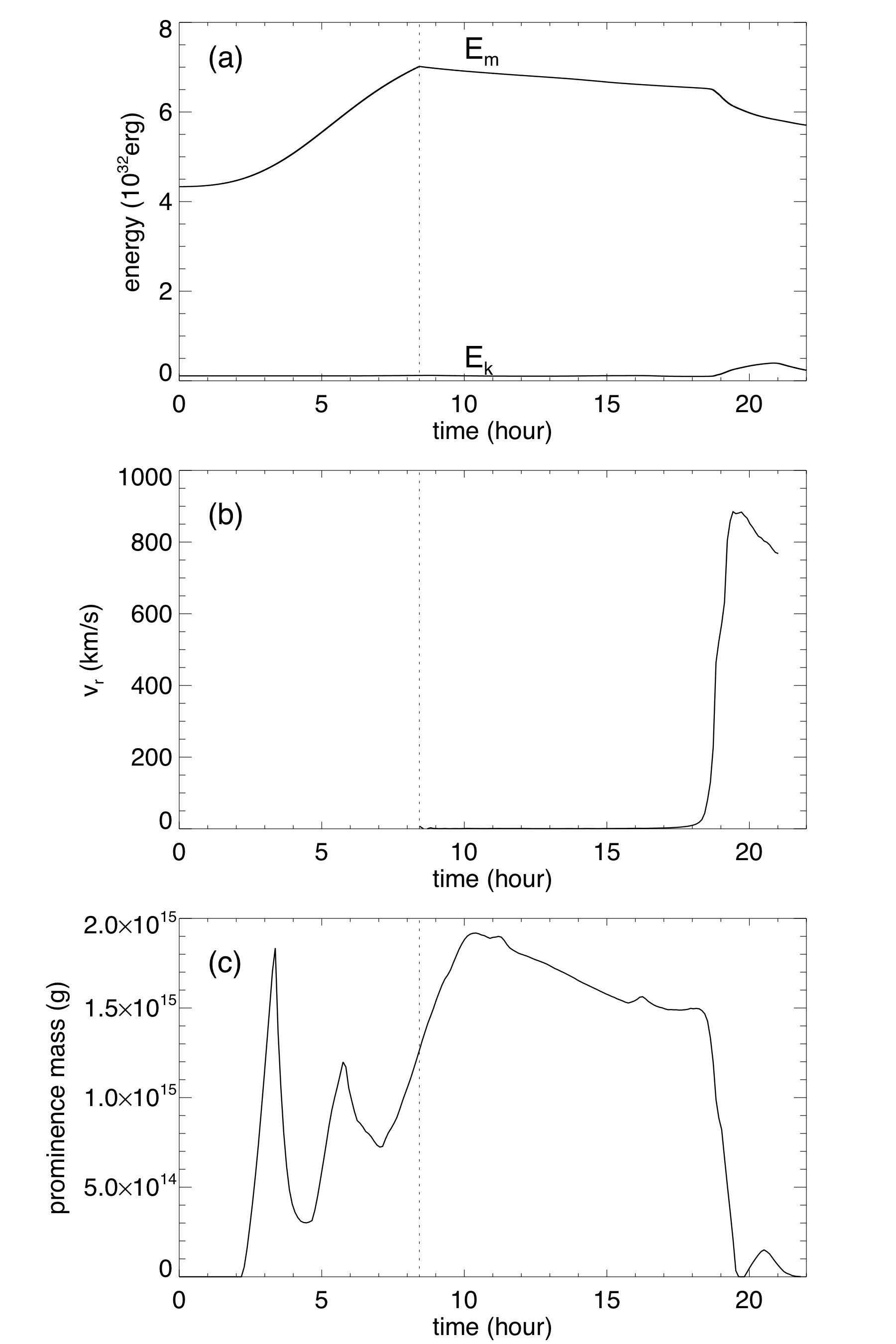}% This is a *.jpg file
\end{center}
\caption{(a) The evolution of the total magnetic energy $E_m$ and total kinetic
energy $E_k$, (b) the evolution of the rise velocity at the apex
of the axial field line of the emerged flux rope, and (c) the evolution
of the cool prominence mass in the corona evaluated as the total mass with temperature
below $10^5$ K.}
\label{fig:emekvr}
\end{figure}

\begin{figure}[ht!]
\begin{center}
\includegraphics[width=4in]{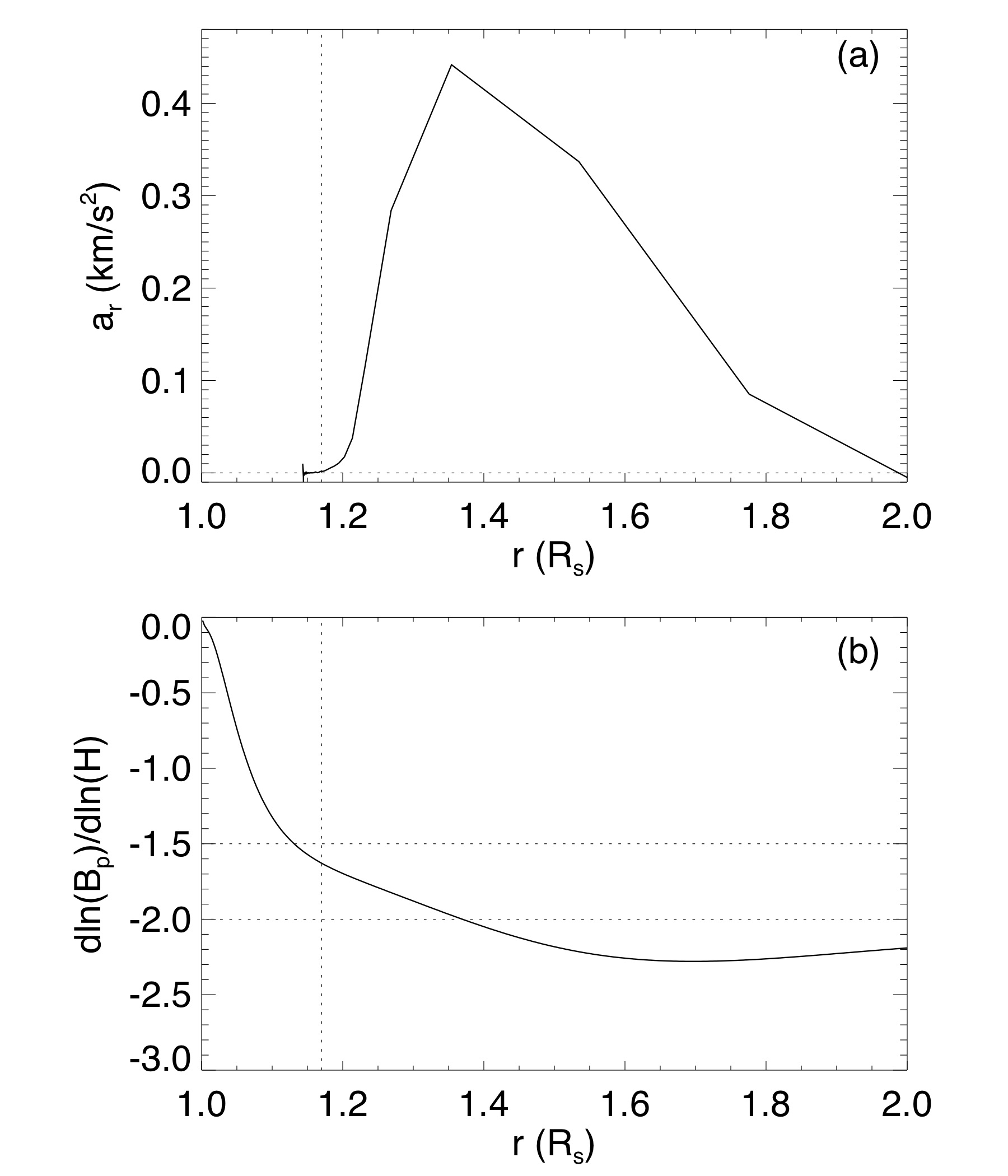}% This is a *.eps file
\end{center}
\caption{(a) Acceleration at the apex of the flux rope's axial field line
as a function of its height position, and (b) the
decay rate of the corresponding potential magnetic
field strength $B_p$ with height $H$ (above the surface) when the emergence is stopped
(after which the lower boundary normal magnetic flux distribution and the
corresponding potential field remain fixed).}
\label{fig:ar_bpdecay}
\end{figure}

Figures \ref{fig:evol_sideview}(a)-(f) show snapshots of the 3D coronal magnetic field
lines during the course of the evolution of the emerged coronal flux rope, from the
quasi-static phase to the onset of eruption.
The field lines shown in the snapshots are selected as follows. A set of field lines
from a set of fixed foot points in the pre-existing bipolar bands are traced as the red
field lines (same field lines as those traced in Figure \ref{fig:init_streamer}(d) for
the initial state).
For the representative field lines in the emerged flux rope, we trace
field lines from a set of tracked foot points at the lower boundary that connect to
a set of selected field lines of the subsurface emerging torus and color the field lines
(green, cyan, and blue) based on the flux surfaces of the subsurface torus.
%An ``axial field line'' (black) is traced from the foot points that connect to
%the axis of the subsurface torus.
Figures \ref{fig:evol_sideview}(g)-(l) show the synthetic SDO/AIA EUV
images in 304 {\AA} channel as viewed from the same line of sight (LOS)
corresponding to the snapshots shown in Figures \ref{fig:evol_sideview}(a)-(f).
The synthetic AIA images are computed by integrating along individual
line-of-sight (LOS) through the simulation domain:
\begin{equation}
I_{\rm channel} = \int n_e^2 (l) \, f_{\rm channel} (T(l)) \, dl,
\label{eq:I_channel}
\end{equation}
where $l$ denotes the length along the LOS through the simulation
domain, $I_{\rm channel}$ denotes the integrated emission intensity at
each pixel of the image in
units of DN/s/pixel (shown in LOG scale in the images),
``channel'' denotes the AIA wavelength channel
(which is 304 {\AA} in the case for Figures \ref{fig:evol_sideview}(g)-(l)),
$n_e$ is the electron number density, and
$f_{\rm channel} (T)$ is the temperature response function that takes into
account the atomic physics and the properties of the AIA channel filter.
We obtain the temperature dependent function $f_{\rm channel} (T)$ for
the individual filters using
the SolarSoft routine {\tt get\_aia\_response.pro}.
The response function for the AIA 304 {\AA} channel peaks at the
temperature of about $8 \times 10^4 $K, thus the synthetic emission
images show where the cool prominence plasma condensations form in
the flux rope.
For the LOS integration, we also have assumed that the prominence
condensations are ``optically thick'' such that when the LOS reaches
a plasma where both the temperature goes below $7.5 \times 10^4$ K and
the number density is above $10^9 {\rm cm}^3$, we stop the integration for
that LOS assuming the emission from behind the plasma is blocked and
does not contribute to the integrated emission for the LOS.
Figure \ref{fig:emekvr} shows the evolution of the total magnetic energy $E_m$,
the total kinetic energy $E_k$, the rise velocity at the apex of the
axial field line of the emerged flux rope, and the temporal evolution of
the cool prominence mass in the corona evaluated as the total mass with
temperature below $10^5$ K.
From $t=0$ to $8.42$ hour, $E_m$ increases as the emergence
of a twisted magnetic torus is imposed at the lower boundary, and a long
coronal flux rope is built up quasi-statically, confined by the coronal streamer as can be
seen in the snapshot in Figure \ref{fig:evol_sideview}(a) at $t=8.42$ hour.
The emergence is stopped at $t=8.42$ hour at which time the total field line
twist about the axial field line of the emerged flux rope reaches about
$1.76$ winds between the anchored ends. This twist is above the critical value
(about 1.25 winds) for the onset of the kink instability for a simple
1-dimensional cylindrical line-tied force-free flux rope \citep{Hood:Priest:1981}.
However subsequently, the flux rope is found to settle into a 
quasi-static rise phase over a long period of time (corresponding to
about $131$ Alfv\'en crossing times along the axis), from $t=8.42$ hour
to about $t=17$ hour, with nearly zero acceleration (Figure \ref{fig:emekvr}(b)).
We find that a long extended prominence has formed in the emerged flux rope
(see Figures \ref{fig:evol_sideview}(g)(h)),
with the prominence condensations in the dips of the twisted field
lines. In the PROM simulation in F18, it is shown
that the cool prominence condensations form
due to the development of the radiative instability of the dense plasma
in the dips after their emergence.
Here in the present case we find that cool prominence condensations
begin to form even earlier in the flux emergence,
soon after the apex of the flux rope emerges, as dense plasma is pushed into
the corona with the stronger flux rope field.
However these earlier forming condensations are unsteady and drain down
as the flux rope emergence continues, and later more stable condensations
form in the dips of the emerged field lines as in the PROM case.
In Figure \ref{fig:emekvr}(c) we see large temporal fluctuations of the cool
prominence mass during the early phase of the flux emergence.
After the emergence is stopped (marked by the vertical dotted line in Figure
\ref{fig:emekvr}),
the prominence mass shows both a phase of continued increase and
then a gradual decrease during the quasi-static phase
(from $t=8.42$ hour to about $t=17$ hour).
However this change from an increase of prominence mass to
a decline does not seem to be associated with any significant change in the rise
velocity in the quasi-static phase.
During this quasi-static rise period the magnetic energy decreases slowly
(see Figure \ref{fig:emekvr}) due to the continued magnetic reconnections,
until about $t=17$ hour when
it reaches a height of about $r=1.17 R_s$, where the flux rope
begins a monotonic acceleration (see Figure \ref{fig:ar_bpdecay}(a))
and erupts subsequently with a sharp decrease of the magnetic energy and
a significant increase of the kinetic energy (see Figure \ref{fig:emekvr}(a)).
Figure \ref{fig:ar_bpdecay} shows that the height (marked by the vertical dotted line)
at which the flux rope begins
to accelerate monotonically (see panel (a)) corresponds to the height at which
the decay rate of the corresponding potential field reaches a magnitude of
about  1.6 (see panel (b)), which exceeds the critical decline rate of
about 1.5 for the onset of the torus instability for a toroidal flux
rope \citep[e.g.][]{Kliem:Toeroek:2006}.
Thus in the present case, the onset of eruption is compatible with the onset
of the torus instability.
We find that in the present simulation, the flux rope begins to erupt at 
a significantly lower height (at about $r=1.17 R_s$) compared to
that (at about $r=1.6 R_s$) in the PROM simulation in F18.
This is because the arcade field lines in the pre-existing streamer of the
PROM simulation have a significantly wider mean foot-point separation and hence
the corresponding potential field declines with height significantly more
slowly, where the magnitude of the decay rate remains below 1.5 until about
$r=1.3 R_s$ in that case.

\begin{figure}[ht!]
\begin{center}
\includegraphics[width=4in]{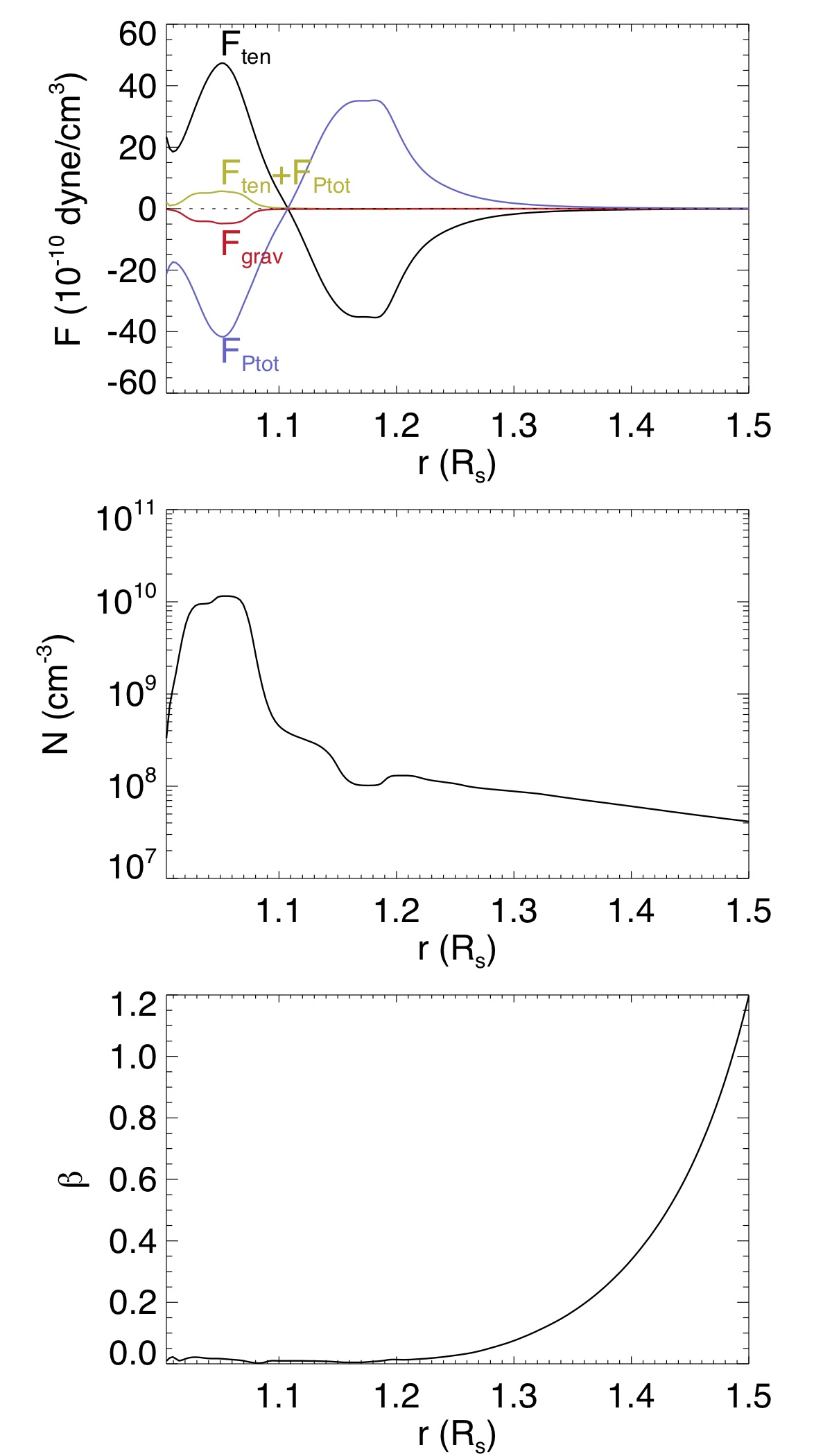}% This is a *.jpg file
\end{center}
\caption{Several radial forces (top), density (middle), and plasma-$\beta$
(the ratio of gas pressure to magnetic pressure) (bottom) along the central vertical
line through the middle of the flux rope in Figure \ref{fig:evol_sideview}(b).
The radial forces shown in the top panel are the magnetic tension force $F_{\rm ten}$
(black curve), the total pressure gradient force $F_{\rm Ptot}$ (blue curve),
which is predominantly the magnetic pressure gradient (because of the low plasma-$\beta$)
as shown in the bottom panel), the sum $F_{\rm ten}+ F_{\rm Ptot}$ (green curve),
which is approximately the net Lorentz force,
and the gravity force of the plasma $F_{\rm grav}$ (red curve).
as a function of height.}
\label{fig:forces}
\end{figure}

Although the confining potential field in the present case declines with
height more steeply, it is stronger lower down and hence the flux rope and
the prominence during the quasi-static phase reach significantly lower heights
compared to the PROM case, and the flux rope field strength is also
significantly stronger.
The peak Alfv\'en speed in the central flux rope cross-section in the present
case reaches about $4100$ km/s with a peak field strength of
about $24$G, compared to the peak Alfv\'en speed of about $1500$ km/s and
peak field strength of about $9$G in the PROM case during the quasi-static
stage. The stronger flux rope field strength causes the prominence-carrying field
to be much closer to force-free compared to the PROM case, as shown in
Figure \ref{fig:forces} compared to Figure 6 in F18.
Figure \ref{fig:forces} shows that the net Lorentz force (the green curve in
the top panel) that balances the
gravity force (the red curve) of the prominence is at most about 0.1 of
either the magnetic tension or magnetic pressure gradient. In contrast in the
PROM case in F18, there is a significant net Lorentz force to balance the prominence
gravity that is comparable to the magnetic tension, i.e. the prominence carrying
fields in the flux rope is significantly non-force-free. 
In the present case however, the magnetic field is close to force-free throughout
the flux rope, even for the prominence-carrying field.
Thus we do not see a significant variation of the rise velocity during the
quasi-static phase in response to the growth or decline of prominence condensation
mass as found above, and the onset of eruption is consistent with the onset
of the torus instability.
The stronger flux rope field strength in the present case also produces
a stronger acceleration and a higher peak velocity of the erupting
flux rope.
In the present case the flux rope is found to
accelerate to a peak velocity of nearly $900$ km/s (see Figure \ref{fig:emekvr}(b)),
compared to the peak velocity of about $600$ km/s reached in the PROM
simulation (see Figure 4(b) in F18).
However, the ratio of the peak velocity over the peak Alfv\'en speed of the
flux rope is found to be lower (0.22) in the present case compared to the PROM
case (0.4).

\begin{figure}[ht!]
\begin{center}
\includegraphics[width=6in]{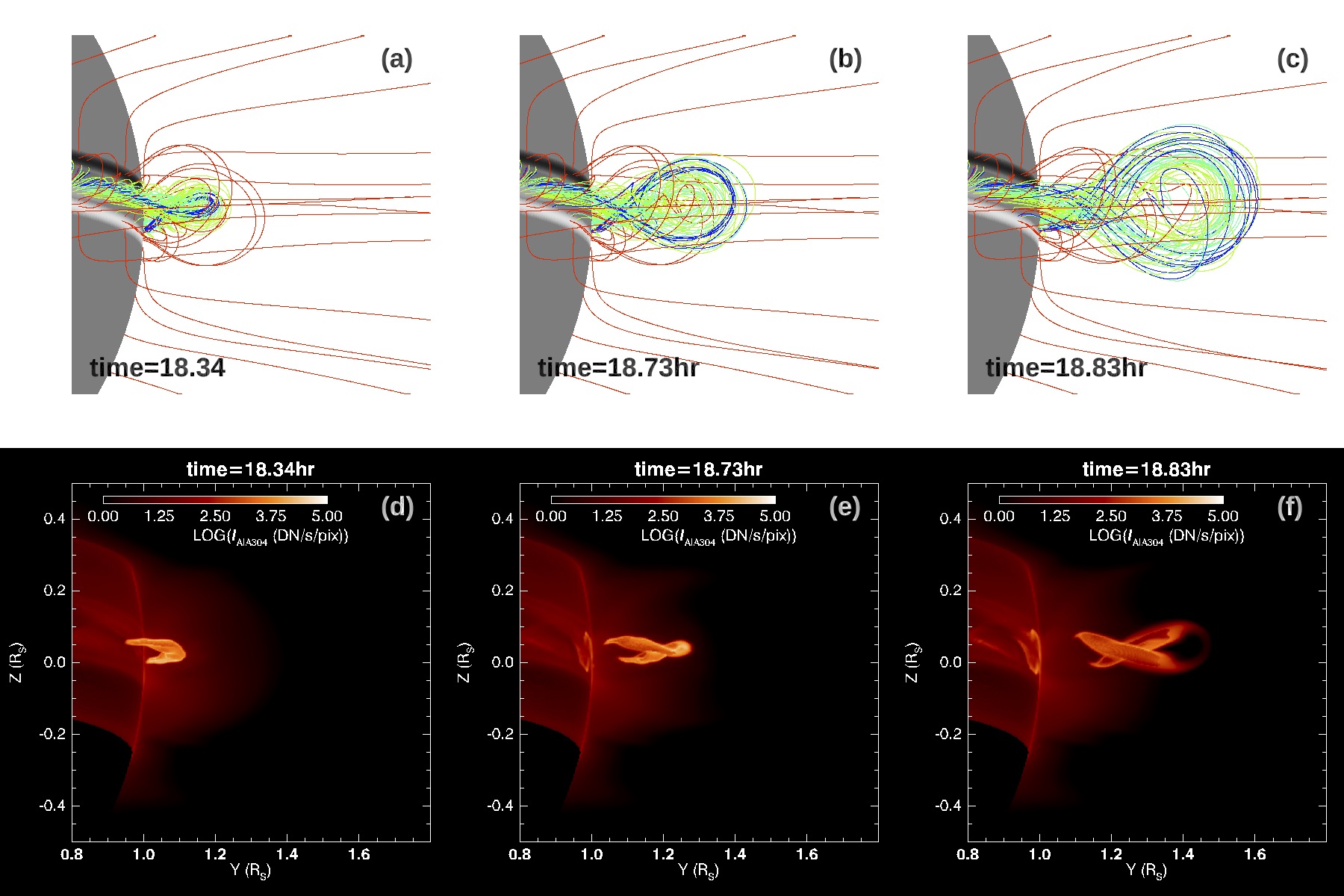}% This is a *.jpg file
\end{center}
\caption{Successive snapshots of the erupting flux rope field lines ((a)(b)(c))
and the corresponding synthetic AIA 304 {\AA} images ((d)(e)(f)), viewed
with a LOS that is close to aligned with the length of the flux rope.
They show the writhing motion of the flux rope and the associated
erupting prominence.}
\label{fig:evol_limbview}
\end{figure}

Although the onset of eruption in the present simulation is consistent
with the onset of the torus instability, because of the significant total
twist in the emerged flux rope, the erupting flux rope
shows significant rotational motion and a kinked morphology as can be seen
in Figures \ref{fig:evol_sideview}(d)-(f).
The associated erupting prominence also shows a kinked morphology (see
Figures \ref{fig:evol_sideview}(j)-(l)).
The rotation of the erupting prominence is more clearly seen from the
view shown in Figure \ref{fig:evol_limbview}, where the flux rope is
viewed nearly along its length.
We can clearly see the writhing motion of the erupting prominence
due to the writhing motion of the hosting flux rope.

We also note that two brightening ribbons are visible in the AIA 304 {\AA} images
(Figures \ref{fig:evol_limbview}(e)(f)) on the lower boundary under the erupting prominence.  The
brightening ribbons correspond to the foot points of the highly heated, post-reconnection
loops just reconnected in the flare current sheet behind the erupting flux rope.
The strong heat conduction flux coming down along the heated post-reconnection
loops causes an increase of the pressure and density at the foot points at the
lower boundary (based on the variable pressure lower boundary condition used here
as described in F17).  This enhanced density at the foot points leads to the
brightening of the ribbons in 304 channel emission. They qualitatively represent
the flare ribbons regularly seen in eruptive flares.

\subsection{The formation of prominence-cavity system}

\begin{figure}[ht!]
\begin{center}
\includegraphics[width=6in]{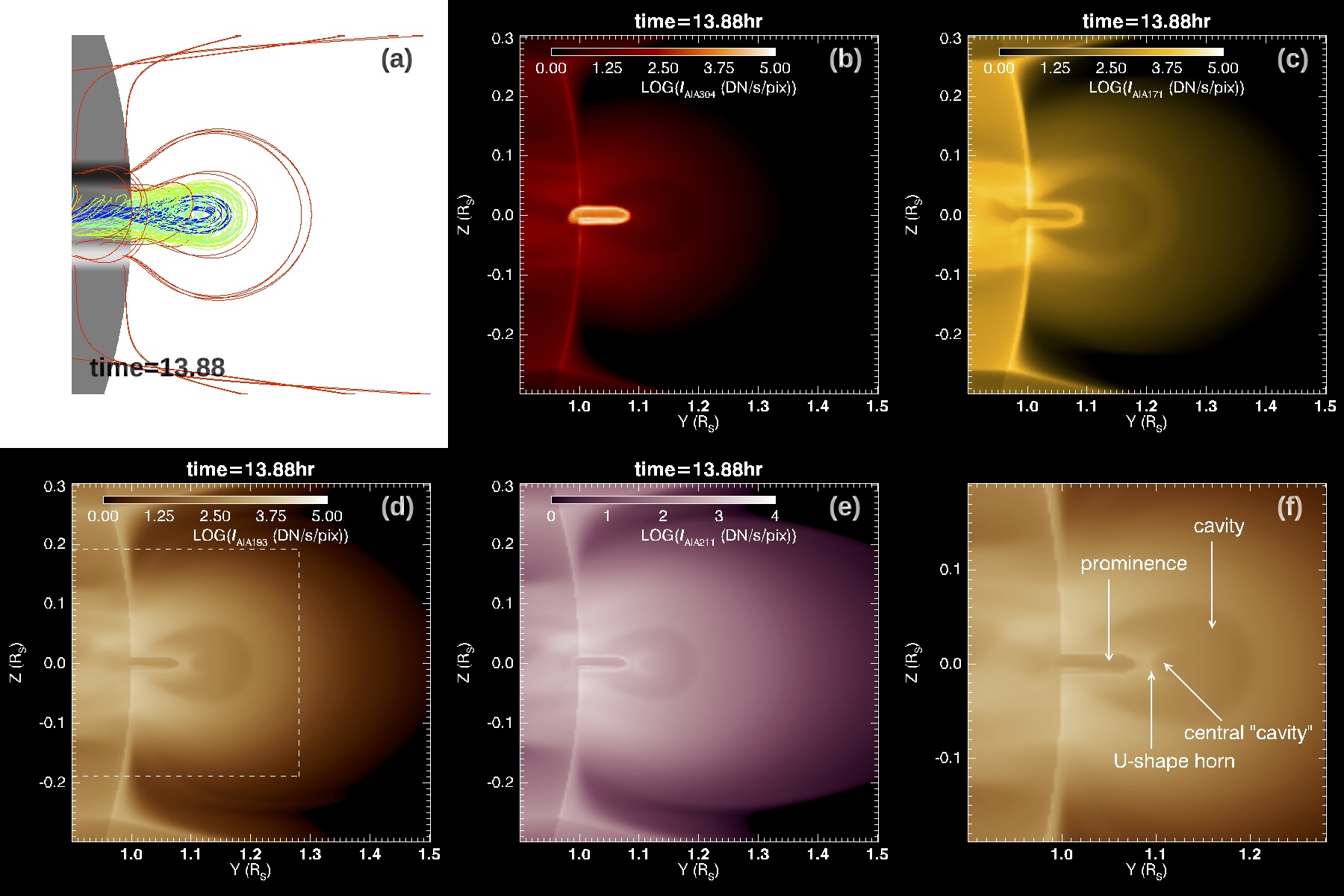}% This is a *.jpg file
\end{center}
\caption{3D field lines (a), and synthetic SDO/AIA EUV
images in 304 {\AA} (b), 171 {\AA} (c), 193 {\AA} (d), and 211 {\AA} (e) channels,
with the flux rope viewed along its axis above the limb, at time $t=13.88 {\rm hr}$
during the quasi-static phase. Panel (f) shows the zoomed in view of the boxed area of panel
(d) with the cavity substructures labeled.}
\label{fig:euv_it140}
\end{figure}

\begin{figure}[ht!]
\begin{center}
\includegraphics[width=6in]{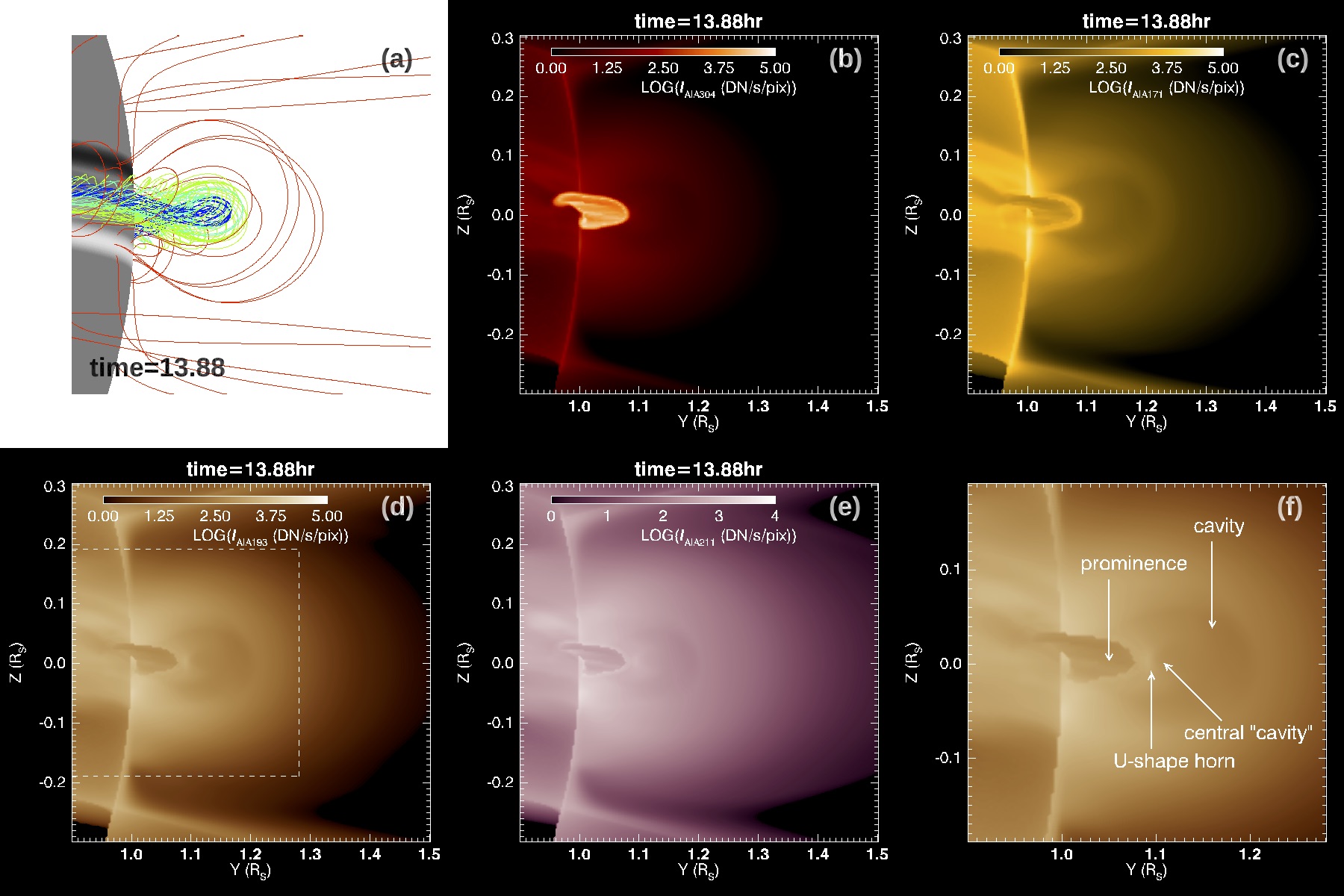}% This is a *.jpg file
\end{center}
\caption{Same as Figure \ref{fig:euv_it140} but with the flux rope viewed slightly tilted
by $5^{\circ}$.}
\label{fig:euv_tilt_it140}
\end{figure}

Figure \ref{fig:euv_it140} shows the limb view of the 3D magnetic field lines (panel (a)),
and synthetic SDO/AIA EUV images in 304 {\AA} (panel (b)), 171 {\AA} (panel (c)),
193 {\AA} (panel (d)), and 211 {\AA} (panel (e)) channels,
with the flux rope viewed along its axis at a time ($t=13.88 {\rm hr}$)
during the quasi-static phase.
A similar limb view with the flux rope slightly tilted by $5^{\circ}$
is shown in Figure \ref{fig:euv_tilt_it140}.
The AIA 171 {\AA}, 193 {\AA}, 211 {\AA} channel images are computed in the same way as
described in the previous section using equation \ref{eq:I_channel} with the temperature
response function $f_{\rm channel} (T)$ replaced by that for the corresponding channel.
The synthetic AIA images with the flux rope viewed nearly along
its axis (as illustrated in Figures \ref{fig:euv_it140}
and \ref{fig:euv_tilt_it140})
show the formation of a prominence-cavity system
with qualitative features similar to observations
\citep[e.g.][]{Gibson:2015}.
Inside the bright helmet dome, we see a dark cavity surrounding
the lower central prominence (which appears dark in the
171 {\AA}, 193 {\AA}, or 211 {\AA} images due to the optically thick
assumption).  In this simulation where we have used a pre-existing
streamer solution with a narrower mean foot-point separation
for the closed arcade field, we obtained a significantly smaller
cavity with lower heights for the cavity (about 0.2 $R_s$) and the
prominence (about 0.1 $R_s$) compared
to the previous PROM simulation in F18
(about 0.47 $R_s$ for the cavity height and 0.17 $R_s$ for the
prominence height),
in better agreement with observations \citep[e.g.][]{Gibson:2015},
which find a median height for EUV cavities of 0.2 $R_s$.
In the synthetic 193 {\AA}, or 211 {\AA} images in
Figures \ref{fig:euv_it140}(d)(e)
and \ref{fig:euv_tilt_it140}(d)(e), we also find substructure inside
the cavity, similar to some of the features described
in \citep[e.g.][]{Gibson:2015,Su:etal:2015}.
We find a central smaller cavity on top of the prominence enclosed by
a``U''-shaped or horn-like bright structure 
extending above the prominence.  Such substructure is similar
to the features as shown in Figure 8 and Figure 12 in
\citep{Gibson:2015} and Figure 2(c) in \citep{Su:etal:2015}.
Here we examine the characteristics of the 3D magnetic field comprising
the different parts of the prominence-cavity system formed in our
MHD model.

\begin{figure}[ht!]
\begin{center}
\includegraphics[width=6in]{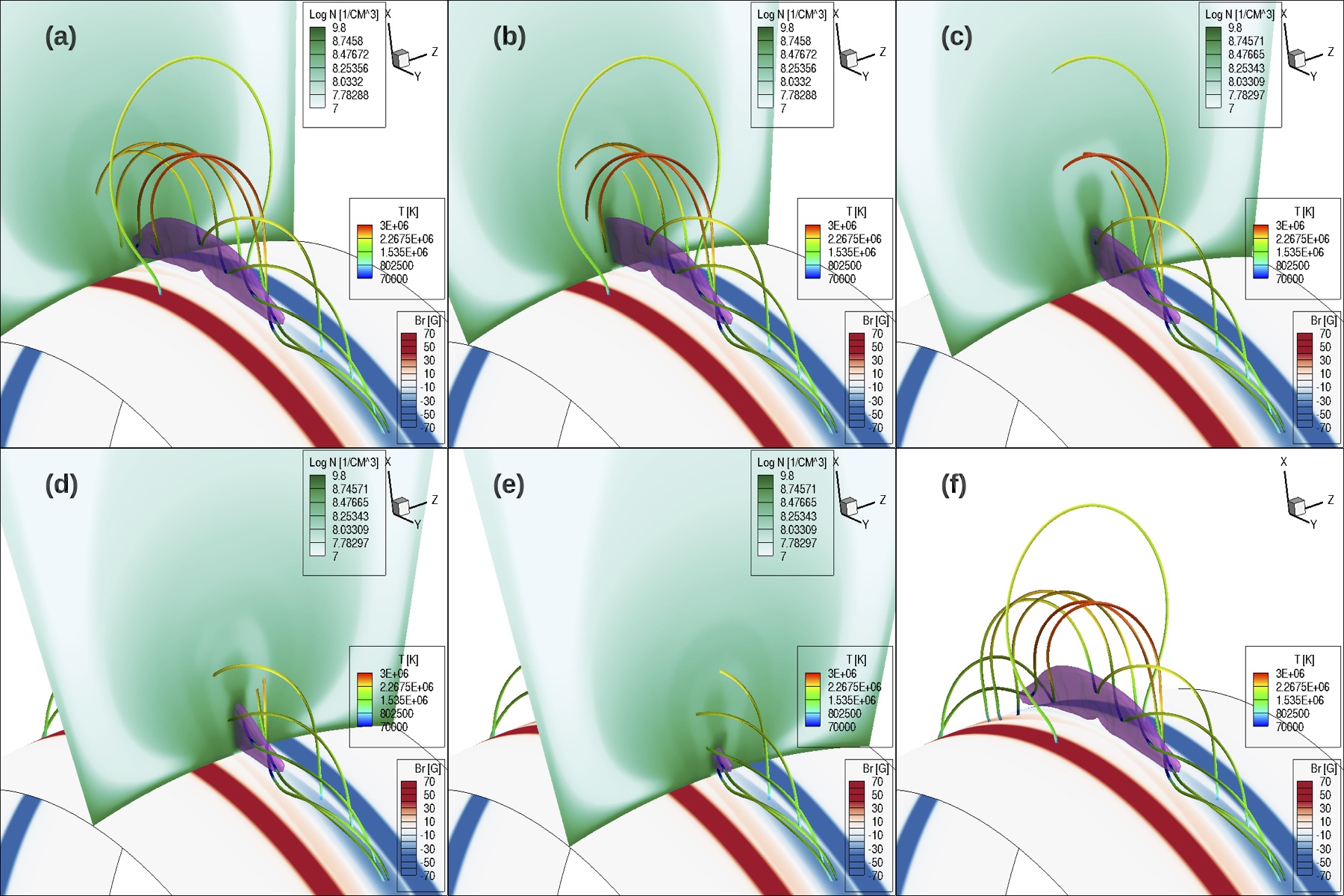}% This is a *.jpg file
\end{center}
\caption{A set of prominence-carrying field lines containing prominence dips and one
representative arcade dome field line, all colored with temperature, plotted with a
cross-section of density placed at different locations along the flux rope for the
different panels (a)-(e).  Panel (f) shows the same field lines 
without the density cross-section. A pink iso-surface of temperature at
$T=7.5 \times 10^4 $K outlines the location of the prominence condensation.
The lower boundary surface is colored with the normal magnetic field strength.
All the images are at the time $t=13.88 {\rm hr}$ during the quasi-static stage.}
\label{fig:cavity_fdls}
\end{figure}

\begin{figure}[ht!]
\begin{center}
\includegraphics[width=6in]{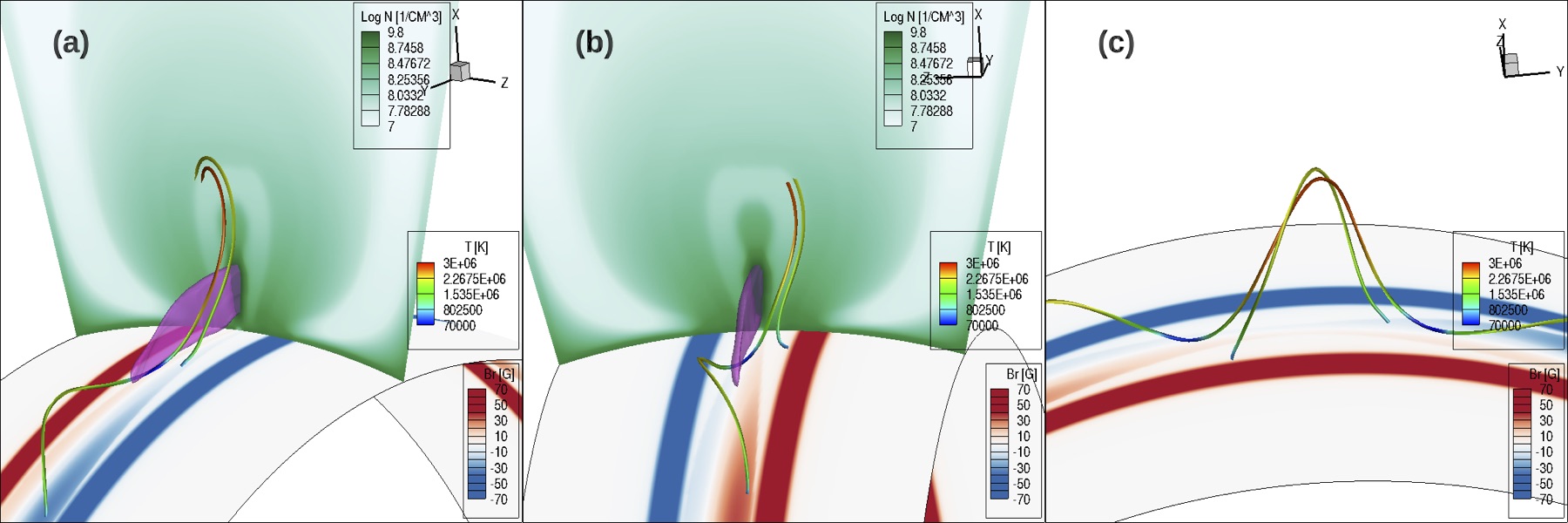}% This is a *.jpg file
\end{center}
\caption{Two field lines traced from two adjacent points on the two sides of
the cavity boundary in the density cross-section shown in (a) and (b) viewed
from two different perspectives from opposite sides of the cross-section.
The pink iso-surface of temperature at $T=7.5 \times 10^4$ K outlines the prominence
condensations. (c) shows the same field lines from a different view without
the density cross-section and the iso-surface for the prominence. The lower boundary
shows the normal magnetic field distribution $B_r$.}
\label{fig:cavity_boundary}
\end{figure}

Figure \ref{fig:cavity_fdls} shows a set of selected
prominence-carrying field lines that contain prominence dips
and one representative arcade dome field line in the high density
dome, together with a vertical cross-section of density placed
at different locations (for the different panels (a)-(e))
along the flux rope.
The prominence condensation is outlined
by the pink temperature iso-surface with $T=7.5 \times 10^4$K.
It can be seen in Figures \ref{fig:cavity_fdls}(a)-(e) that
as the density cross-section slides along the flux rope, the prominence
carrying field lines intersect the cross-section in the low density cavity
region, except at the prominence dips.
In other words, we find that the prominence and the surrounding cavity are
threaded by the prominence-carrying field lines, with the cavity corresponding
to the density-depleted portions of the prominence-carrying field lines
extending up from the prominence dips.
As was shown in F18, the runaway radiative cooling of the prominence
condensations in the dips causes a lowered pressure and draining of plasma
towards the dips, establishing a more rarefied atmosphere along the 
dip-to-apex portions of the prominence carrying field lines compared
to the surrounding dome field lines without dips.
We find that the cavity boundary corresponds to a sharp transition
from the dipped prominence carrying field lines inside the cavity to the
arcade-like field lines without dips outside in the higher density dome.
This is illustrated in an example shown in Figure \ref{fig:cavity_boundary}.
Two field lines are traced from two adjacent points on the two sides of
the cavity boundary, and they show very different connectivity to the lower boundary,
with the one from inside the cavity being a long twisting field line carrying
two prominence dips and the other from just outside the cavity being a significantly
shorter arcade-like field line with no dips (see Figure \ref{fig:cavity_boundary}(c)). 
Because of the drastic difference in the connectivity and
whether there are prominence condensations, the two field lines show very different
thermodynamic properties at their two adjacent points near the cavity boundary,
giving rise to the sharp appearance of the EUV cavity boundary.
Note that the arcade-like field line in the dome region shows mixed types of
foot points, with one foot point connecting to the pre-existing bipolar bands and 
the other foot point in the emerging flux rope foot points, suggesting that there
have been continued
reconnections between the flux rope and the
pre-existing arcade field.

\begin{figure}[ht!]
\begin{center}
\includegraphics[width=6in]{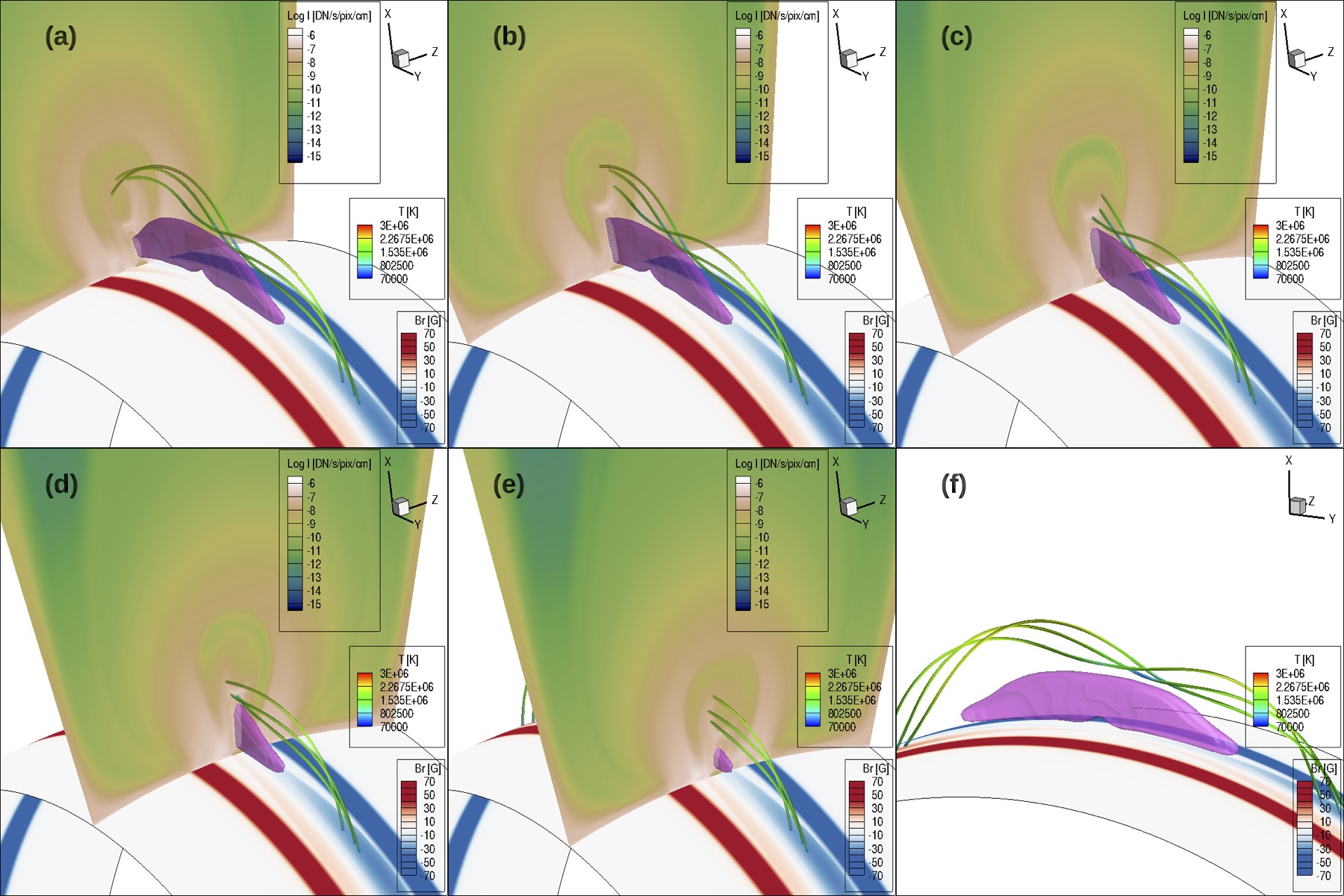}% This is a *.eps file
\end{center}
\caption{A set of field lines threading through the region that contributes to the
EUV 193 {\AA} channel emission that produces the horn-like structure inside the cavity,
together with a cross-section showing the local emission intensity in EUV 193 {\AA} channel
(the integrand ${n_e}^2 f_{193} (T)$ in equation \ref{eq:I_channel}) placed at different
locations along the flux rope for the different panels (a)-(e).
Panel (f) shows the same field lines
without the cross-section and viewed from a different perspective.
The field lines are colored in temperature.
A pink iso-surface of temperature at
$T=7.5 \times 10^4 $K outlines the location of the prominence condensation.
The lower boundary surface is colored with the normal magnetic field strength.
All the images are at the same time ($t=13.88 {\rm hr}$) as those in
Figure \ref{fig:cavity_fdls}.}
\label{fig:horn_fdls}
\end{figure}

\begin{figure}[ht!]
\begin{center}
\includegraphics[width=6in]{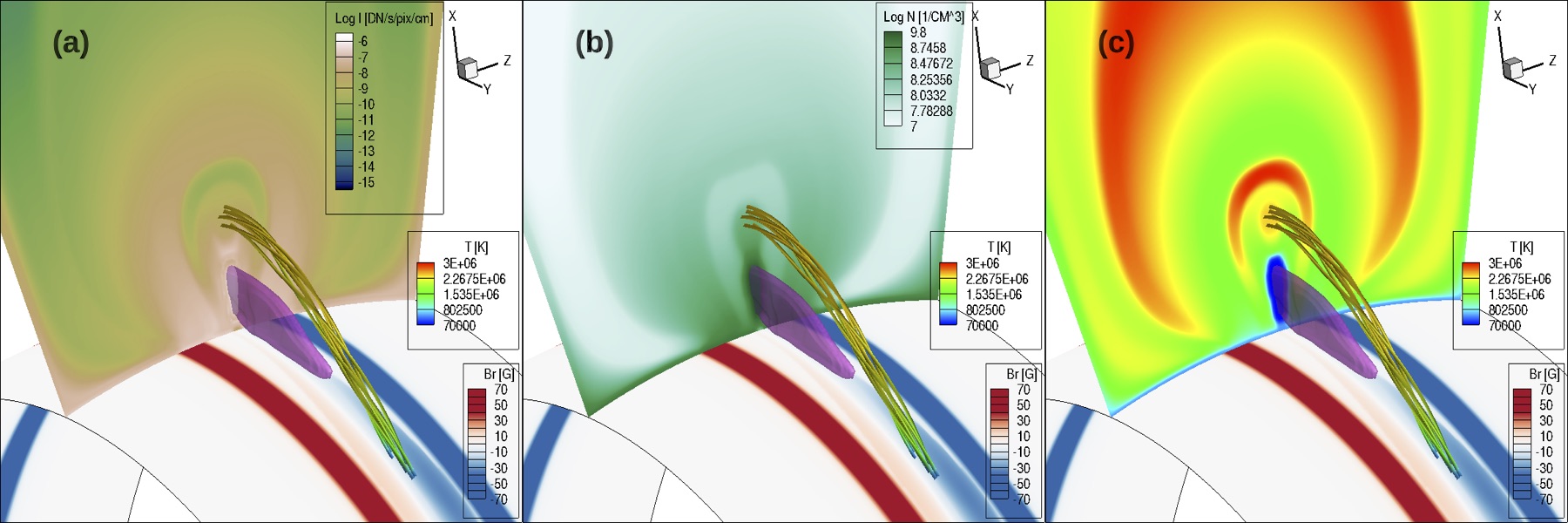}% This is a *.jpg file
\end{center}
\caption{A set of field lines threading through the central EUV cavity enclosed
in the prominence horns, together with a middle cross-section showing the 
distribution of EUV 193 {\AA} channel emission intensity (a), density (b), and
temperature (c). The field lines are colored in temperature and a pink iso-surface
of temperature at $T=7.5 \times 10^4 $K outlining the location of the prominence
condensation is also shown.}
\label{fig:core_fdls}
\end{figure}

To examine the magnetic field that produces the substructure inside the EUV cavity,
we have also traced field lines that thread through the region that contributes to
the EUV emission of the prominence ``horns''.
Figure \ref{fig:horn_fdls} shows a set of such field lines, together with a cross
section showing the local emission intensity in EUV 193 {\AA} channel
(the integrand ${n_e}^2 f_{193} (T)$ in equation \ref{eq:I_channel}), with the
cross-section placed at different locations along the flux rope for the
different panels (a)-(e), and without showing the cross-section in panel (f).
It can be seen that as the cross-section slides along the flux rope, the
field lines intersect the central ``U''-shaped region of enhanced
EUV emission.
As shown in Figure \ref{fig:horn_fdls}(f), we find that these field
lines that contribute to the prominence-horn
emission are field lines containing relatively shallow dips, where the 
prominence condensations have evaporated to coronal temperatures (above
$4 \times 10^5$ K and with most parts of the field lines ranging between
$8 \times 10^5$ K and $2.2 \times 10^6$ K) while the density is still
relatively high compared to the surrounding cavity, and hence producing
a favorable conditions for the enhanced EUV 193 {\AA} channel emission.
The cross-sections showing the EUV 193 {\AA} channel
emission intensity in Figures \ref{fig:horn_fdls}(a)-(e)
also illustrate that enclosed inside the bright ``U''-shaped
prominence ``horns'' is another central region of reduced emission,
corresponding to the central ``cavity'' seen in
the synthetic EUV images
(Figures \ref{fig:euv_it140}(d) \ref{fig:euv_tilt_it140}(d)).
Tracing field lines through this central ``cavity" region, we find that
it is threaded by long twisted field lines that contain no dips as shown
in Figure \ref{fig:core_fdls}.
However as shown in the cross-sections in Figures \ref{fig:core_fdls}(a)(b)(c),
this central ``cavity'' region is rather a relatively dense and hot central core.
Its reduced EUV emission (compared to the horns and the outer
dome region) is due to the high temperature (reaching about 2.5 MK) that
is out of the peak of the EUV response function, instead of due to a low
density as is the case for the outer cavity.
We find that the outer boundary
of the outer cavity has a hot rim of even higher temperature
(see the cross section in Figure \ref{fig:core_fdls}(c)) due to the
heating resulting from continued reconnections between the dipped, twisted
field lines that approach the cavity boundary and their neighboring arcade
like field lines.

\section{Conclusions}
We have carried out MHD simulation of the quasi-static evolution
and onset of eruption of a prominence-forming
coronal flux rope under a coronal streamer, extending the previous work
of F17 and F18.
Previous simulations of the prominence-hosting coronal flux rope
(the WS-L simulation in F17 and the PROM simulation in F18) have
used the WS streamer solution in F17 for the pre-existing
field, whose arcade field lines have a wide mean foot-point separation.
This results in a corresponding potential field that declines slowly
with height.  Consequently the emerged flux rope and the prominence
and cavity that form during the quasi-static stage reach large
heights (larger than typically observed) before the onset of eruption.
For the present simulation we have used a pre-existing streamer
solution with a significantly narrower mean foot-point separation and
stronger foot-point field strength for the arcade field lines.
This results in a stronger field strength
lower down and steeper decline of field strength with height for
the corresponding potential field at the end of the flux emergence.
We still drive the emergence of a similar long twisted flux rope
into the corona as in the PROM case in F18.  Similar to the PROM
case, we find the formation of a prominence-cavity system during
the quasi-static evolution, but with significantly lower heights for
the prominence (reaching about $0.1 R_s$) and the cavity (extend
to about $0.2 R_s$), in better agreement with the properties of
the typically observed quiescent prominence-cavity systems.
We also find the formation of cavity substructures, such as the
prominence ``horns'' and central ``voids'' or ``cavities'' on
top of the prominences, in qualitative agreement with the
observed features \citep[e.g.][]{Gibson:2015,Su:etal:2015}.

We have examined the properties of the magnetic fields 
that comprise the different parts of the prominence-cavity
system seen in the synthetic EUV images from our MHD model to
understand the nature of the corresponding observed features.
We find that the prominence and the outer cavity is composed
of the long twisted field lines with dips that contain
prominence condensations (Figure \ref{fig:cavity_fdls}),
where the cavity is threaded by the density depleted portions
of the field lines extending up from the prominence dips.
As was shown in F18, the formation of the prominence
condensations due to runaway radiative cooling causes an
overall lowered pressure in the dips and plasma draining
down towards the dips such that a more rarefied atmosphere is
established for the dip-to-apex portions of the field lines,
compared to the surrounding
arcade field lines in the denser helmet dome.
We find that the boundary of the outer cavity corresponds
to a sharp transition of field line connectivity, where
neighboring field lines connect very differently to the
lower boundary, with long twisted dipped field lines just
inside the boundary and simple arcade-like field lines with
no dips just outside (Figure \ref{fig:cavity_boundary}).
The very different thermodynamic properties of the two
types of neighboring field lines give rise to the sharp
appearance of the EUV cavity boundary.  There are also
continued magnetic reconnections at the boundary, causing
a high temperature rim at the outer cavity boundary
(see the temperature cross-section shown in
Figure \ref{fig:core_fdls}(c))
In regard to the cavity substructures,
we find that the region of the central ``U''-shaped
prominence ``horns'' with relatively enhanced EUV emission inside
the cavity are threaded by twisted field lines with relatively
shallow dips, where the prominence condensations have evaporated
to coronal temperatures while the density is still relatively
high compared to the surrounding cavity
(Figure \ref{fig:horn_fdls}).
For the central ``void'' or ``cavity'' enclosed in the
prominence ``horns'' on top of the prominence, we
find that it corresponds to a central high temperature
and high density core threaded by long twisted field lines
with no dips (Figure \ref{fig:core_fdls}).
It appears as a central ``void'' with weakened EUV emission
not because of a lower density, but because it is heated
to a high temperature reaching about $2.5$ MK that is
outside of the peak of the AIA 193 {\AA} channel (and also
the AIA 211 {\AA} channel) temperature
response function.
We find that the central high temperature core is growing
over the course of the quasi-static phase.
The prominence ``horns'' and growth of the central
hot core result from a gradual transition of dipped
prominence carrying field lines to un-dipped but still twisted
field lines as they rise quasi-statically with the
dips becoming shallower and the prominence condensations evaporating.
The continued magnetic reconnection at the cavity boundary
between the dipped twisted field lines and their neighboring
arcade like field lines may be contributing to the
quasi-static rise by removing the confining field.
We defer to a follow-up paper to conduct
a quasi-separatrix layer analysis \citep[e.g.][]{Pariat:Demoulin:2012}
to study the 
evolution of magnetic reconnection and how it contributes
to the removal of the prominence mass and
the rise of the flux rope during the quasi-static phase.

As was noted in F18, previous 3D MHD simulation of prominence
formation in a stable flux rope by \citet{Xia:etal:2014}, which includes
the chromosphere as the lower boundary, has also found
the formation of a prominence-cavity system with similar
internal structures in synthetic EUV images. Their
explanation of the structures obtained in their simulation
is different from that found in our
simulation. They found that the central dark cavity enclosed
by the horns is threaded by two types of field lines: both
the dipped twisted field lines and the arched twisted field lines
with no dips, while the outer cavity is formed by arched twisted field
lines with no dips. The prominence horns are due to the LOS emission from
the prominence-corona transition regions of the prominence loaded
dipped field lines. They found that during prominence-cavity formation,
density depletion occurs not only on prominence-loaded field lines threading
the cavity and prominence where in situ condensation happens
(as is the case in our simulation), but also on prominence-free field lines
due to mass drainage into the chromosphere. We do not find the latter type of
field lines forming the cavity. Our outer main cavity is threaded by the
density depleted portions of the prominence carrying dipped field lines, and
the inner cavity is threaded by arched twisted field lines with no dips,
which are not density depleted but appear dark in the EUV emission because
they are heated to a high temperature (about $2.5$ MK).
In our simulations, the exclusion of
the chromosphere and fixing the lower boundary at the transition region
temperature do not allow modeling the change of the transition region height
and hence limit the ability to model the condensation/drainage of plasma to
the chromosphere with the cool chromosphere temperature region extending upwards.
Furthermore our 3D simulations that model both the quasi-static phase and the
eruption of prominence-carrying coronal flux ropes have much lower numerical
resolution (1.9 Mm), compared to that achieved in
\citet{Xia:etal:2014,Xia:Keppens:2016a}, which use adaptive grid refinement.
As described in F17 we have modified the radiative loss function to
suppress cooling for $T \leq 7 \times 10^4$ K, so that the smallest pressure
scale height (about 4.4 Mm) of the coolest plasma that forms does not go below two
grid points given our simulation resolution.
The low numerical resolution causes large numerical diffusion and viscosity that can
impact significantly the heating and hydrodynamic evolution of the plasma.
Because of the above limitations of our 3D simulations, the results of the thermodynamic
properties of the resulting prominence-cavity system have large uncertainties, and need
to be confirmed or revised by future higher resolution simulations that include the
chromosphere in the lower boundary, which are our future work.
The current simulation qualitatively illustrates the effect of
the runaway radiative cooling of the prominence condensations in the dips of
the twisted field lines that causes drainage of plasma of the upper portions of
these field lines and creates a cavity with a relatively sharp boundary that
corresponds to the 
transition from the dipped prominence carrying field lines to neighboring
arcade-like field lines. It does not explain the formation of filament channels or
coronal cavities in the absence of filament or prominence condensations.

We find that in the current simulation with a significantly
narrower mean foot-point separation for the arcade field of the
pre-existing streamer, the emerged flux rope begins to erupt at
a significantly lower height compared to the PROM case shown
in F18.  This is due to the steeper decline with height
of the corresponding potential field which allows the onset
of the torus instability lower down.
The eruption also produces a significantly faster CME
compared to the PROM case,
mainly due to the stronger field strength of the pre-eruption
flux rope confined lower down.
Although we find that the onset of eruption in the present
case is consistent with the onset of the torus instability, due
to the large total twist (about 1.76 winds of field line twist)
in the emerged flux rope, both the erupting flux rope and the
associated erupting prominence show significant rotational motion
and develop a kinked morphology (Figures \ref{fig:evol_sideview}
and \ref{fig:evol_limbview}).

\section*{Conflict of Interest Statement}

The authors declare that the research was conducted in the absence of any commercial or financial relationships that could be construed as a potential conflict of interest.

\section*{Author Contributions}

YF is the primary author of the paper who carried out the simulation, computed the synthetic
images and directed the data analysis.  TL carried out the 3D analysis of the simulation
data that lead to the major findings of the paper and contributed to the writing of the paper.

\section*{Funding}
NCAR is sponsored by the National Science Foundation. YF is supported in part
by the Air Force Office of Scientific Research grant FA9550-15-1-0030 to NCAR.
Visiting graduate student TL is supported by the scholarship from the
Chinese Scholarship Council of the Ministry of
Education of China, the National Natural Science Foundation of China grant NOS.
11473071 and 11790302 (11790300) and the Natural Science Foundation of Jiangsu
Province (China) grant No. BK20141043.

\section*{Acknowledgments}
We thank Dr. Jie Zhao for reading the paper and helpful comments on the paper.

%%\section*{Supplemental Data}
 %%\href{http://home.frontiersin.org/about/author-guidelines#SupplementaryMaterial}{Supplementary Material} should be uploaded separately on submission, if there are Supplementary Figures, please include the caption in the same file as the figure. LaTeX Supplementary Material templates can be found in the Frontiers LaTeX folder.

%%\section*{Data Availability Statement}
%%The datasets [GENERATED/ANALYZED] for this study can be found in the [NAME OF REPOSITORY] [LINK].
%% Please see the availability of data guidelines for more information, at https://www.frontiersin.org/about/author-guidelines#AvailabilityofData

%%\bibliographystyle{frontiersinSCNS_ENG_HUMS} % for Science, Engineering and Humanities and Social Sciences articles, for Humanities and Social Sciences articles please include page numbers in the in-text citations
%%%\bibliographystyle{frontiersinHLTH&FPHY} % for Health, Physics and Mathematics articles
%%\bibliography{test}

%%% Make sure to upload the bib file along with the tex file and PDF
%%% Please see the test.bib file for some examples of references

\end{document}